\documentclass[conference]{IEEEtran}
\IEEEoverridecommandlockouts
\usepackage{cite}
\usepackage{amsmath,amssymb,amsfonts}
\usepackage{algorithmic}
\usepackage{graphicx}
\usepackage{textcomp}
\usepackage{float}
\usepackage{stfloats}
\usepackage{textcomp}
\usepackage{multirow}
\usepackage{xcolor}
\def\BibTeX{{\rm B\kern-.05em{\sc i\kern-.025em b}\kern-.08em
    T\kern-.1667em\lower.7ex\hbox{E}\kern-.125emX}}
\begin{document}

\title{An Event-Driven Spiking Compute-In-Memory Macro based on SOT-MRAM
}

\author{
Deyang Yu, Chenchen Liu$^{*}$, Chuanjie Zhang, Xiao Fang, Weisheng Zhao\\
\textit{School of Integrated Circuit Science and Engineering, Beihang University, Beijing 100191, China}\\
Email: \{deyangyu, chenchenliu$^{*}$, chuanjiez, xiao\_fang, weisheng.zhao\}@buaa.edu.cn \\
}
\maketitle



\begin{abstract}

The application of Magnetic Random-Access Memory (MRAM) in computing-in-memory (CIM) has gained significant attention. However, existing designs often suffer from high energy consumption due to their reliance on complex analog circuits for computation. 
In this work, we present a Spin-Orbit-Torque MRAM(SOT-MRAM)-based CIM macro that employs an event-driven spiking processing for high energy efficiency.
The SOT-MRAM crossbar adopts a hybrid series-parallel cell structure to efficiently support matrix-vector multiplication (MVM).
Signal information is (en) decoded as spikes using lightweight circuits, eliminating the need for conventional area- and power-intensive analog circuits.
The SOT-MRAM macro is designed and evaluated in 28nm technology, and experimental results show that it achieves a peak energy efficiency of 243.6 TOPS/W, significantly outperforming existing designs.

\end{abstract}

\begin{IEEEkeywords}

Spin-Orbit Torque MRAM, computing-in-memory, spiking processing.

\end{IEEEkeywords}

\noindent\textsuperscript{*}Chenchen Liu is the corresponding author.

\section{Introduction}

In recent years, CIM has emerged as a promising technology to overcome the fundamental memory bandwidth and energy limitations inherent in the traditional von-Neumann architecture~\cite{1,2,3}.
By enabling operations directly within memory arrays, CIM dramatically reduces data movement, thereby improving both computational and energy efficiency~\cite{4,5,6}. 
This architectural innovation has driven the widespread adoption of CIM in accelerating diverse deep neural networks (DNNs)~\cite{7,8,9,10}.
Notably, the recent proliferation of large language models (LLMs)--which are dominated by memory-bound matrix-vector multiplication (MVM) workloads has intensified the demand for efficient CIM solutions~\cite{11}.


CIM architectures have been extensively investigated using non-volatile memory (NVM) technologies such as SRAM, MRAM, ReRAM, \emph{etc.}~\cite{12,13,14}. 
Among them, SOT-MRAM has gained increasing attention due to its high cell density, ultra-fast switching speed, and excellent endurance.
Recent studies have proposed SOT-MRAM crossbar arrays with series- or parallel-connected cell structures to support analog-style MVM~\cite{15,16}.
These designs leverage the intrinsic Ohm's and Kirchhoff's current laws within the crossbar to enable massively parallel analog computation with high throughput and energy efficiency.
However, a major limitation of analog CIM lies in its dependence on power-hungry and area-intensive peripheral analog circuits, such as DACs and ADCs~\cite{16,17}. 
These components introduce substantial system-level overhead, which significantly erodes the energy efficiency and limits the scalability of analog CIM solutions.

To overcome this challenge, spiking-based CIM designs have been proposed, in which the (in) outputs of the crossbar arrays are (en) decoded into spikes ~\cite{18,19,20}.
One of the most straightforward coding schemes is rate coding, where information is represented by the frequency of spikes over a fixed time window. 
Despite its simplicity, rate-coded hardware suffers from low energy efficiency, limited precision, and nonlinear computation.
Temporal coding schemes such as time-to-first-spike (TTFS) and dual-spike coding have been introduced to address these limitations.
These methods encode information in the precise timing between spikes, offering higher information density and fewer spikes per computation.
While promising, their circuit-level implementations still face challenges, including nonlinear signal accumulation and reliance on synchronous global clocks, which hinder the accuracy, energy efficiency, and scalability of the system.

In this work, we propose a novel spiking-based CIM macro built upon SOT-MRAM to address the above challenges in analog CIM design and achieve enhanced computing accuracy and efficiency. 
Specifically, this work has two key contributions:
1) an event-driven spike-based computation scheme that supports digital interfacing and enables energy-efficient in-memory processing;
2) a suite of circuit-level designs optimized for dual-spike encoding, which achieve high linearity and precise temporal computation.
We implement and simulate the SOT-MRAM CIM macro using $128\times128$ SOT-MRAM crossbar arrays under 28nm technology. 
The functionality, computing accuracy, and energy efficiency of the proposed circuits and CIM macro are thoroughly validated. 

\begin{figure}[htbp]
\vspace{-2mm}
\centerline{\includegraphics[width=0.95\columnwidth]{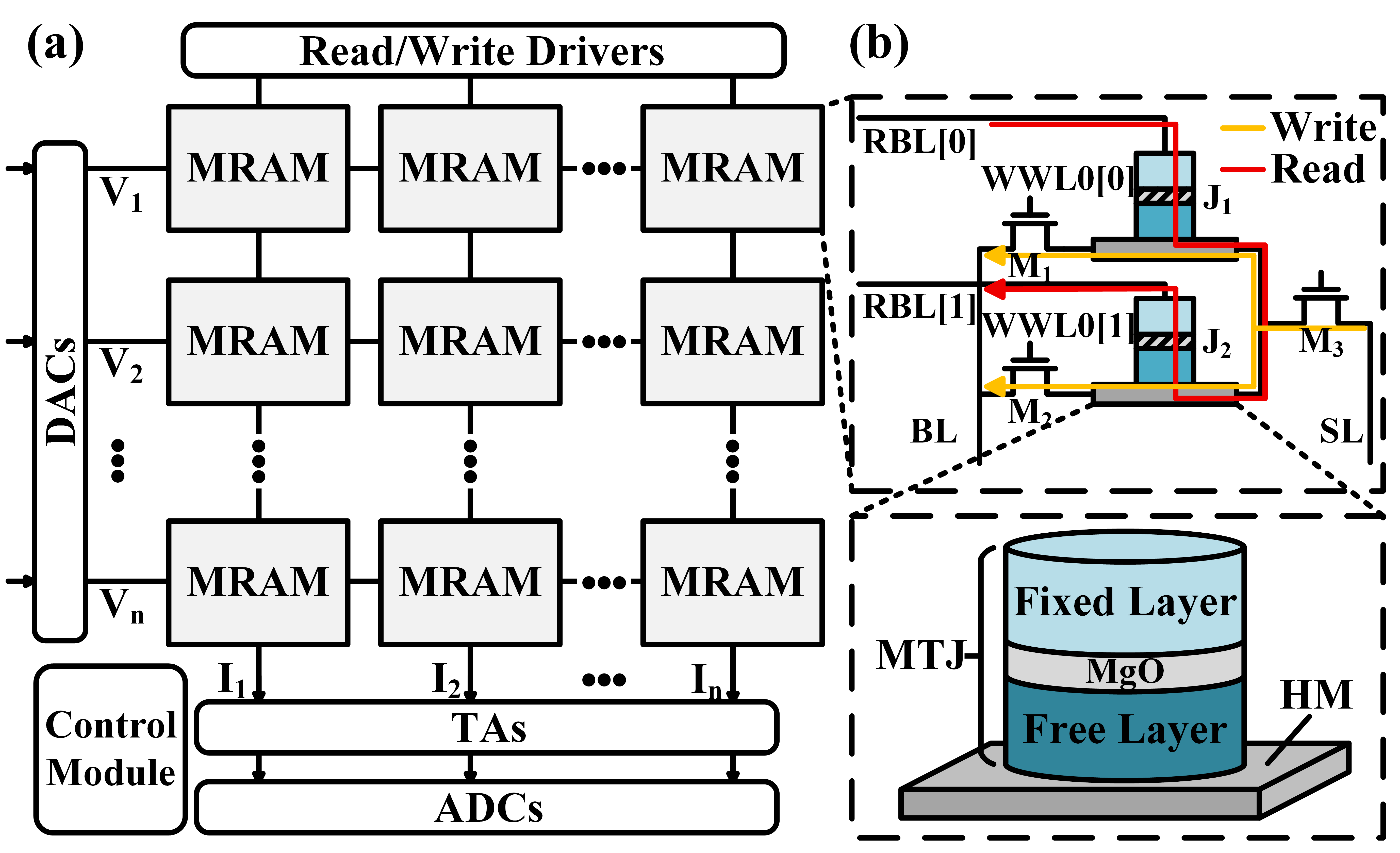}}
\caption{(a) Conventional SOT-MRAM-based CIM designs; (b) Example of a SOT-MRAM cell implementation in 2T-3MTJ structure.}
\label{fig:Traditional}
\end{figure}

\section{Motivation}

\subsection{Conventional SOT-MRAM-based Analog Computing}\label{2A}



Fig.~\ref{fig:Traditional} (a) shows a conventional SOT-MRAM-based CIM architecture.
To perform analog MVM operations, digital inputs are converted to analog voltages via DACs and applied to the word lines of the crossbar array. 
As shown in Fig.~\ref{fig:Traditional} (b), the crossbar is structured in a series-parallel manner, where multiple SOT-MRAM devices are connected in series within each cell. 
This design alleviates the low-resistance issue of MRAM devices and enables multi-bit analog computation~\cite{16}. 
Our proposed architecture also adopts this structure to enhance energy efficiency and scalability. 
During MVM execution, the resulting current is aggregated along each bitline and converted back to digital output using analog peripheral circuits such as TAs and ADCs. 
However, these components are power-hungry and significantly degrade the overall energy efficiency.

\begin{figure}[b]
\begin{center}
\vspace{-2mm}
\includegraphics[width=0.85\columnwidth]{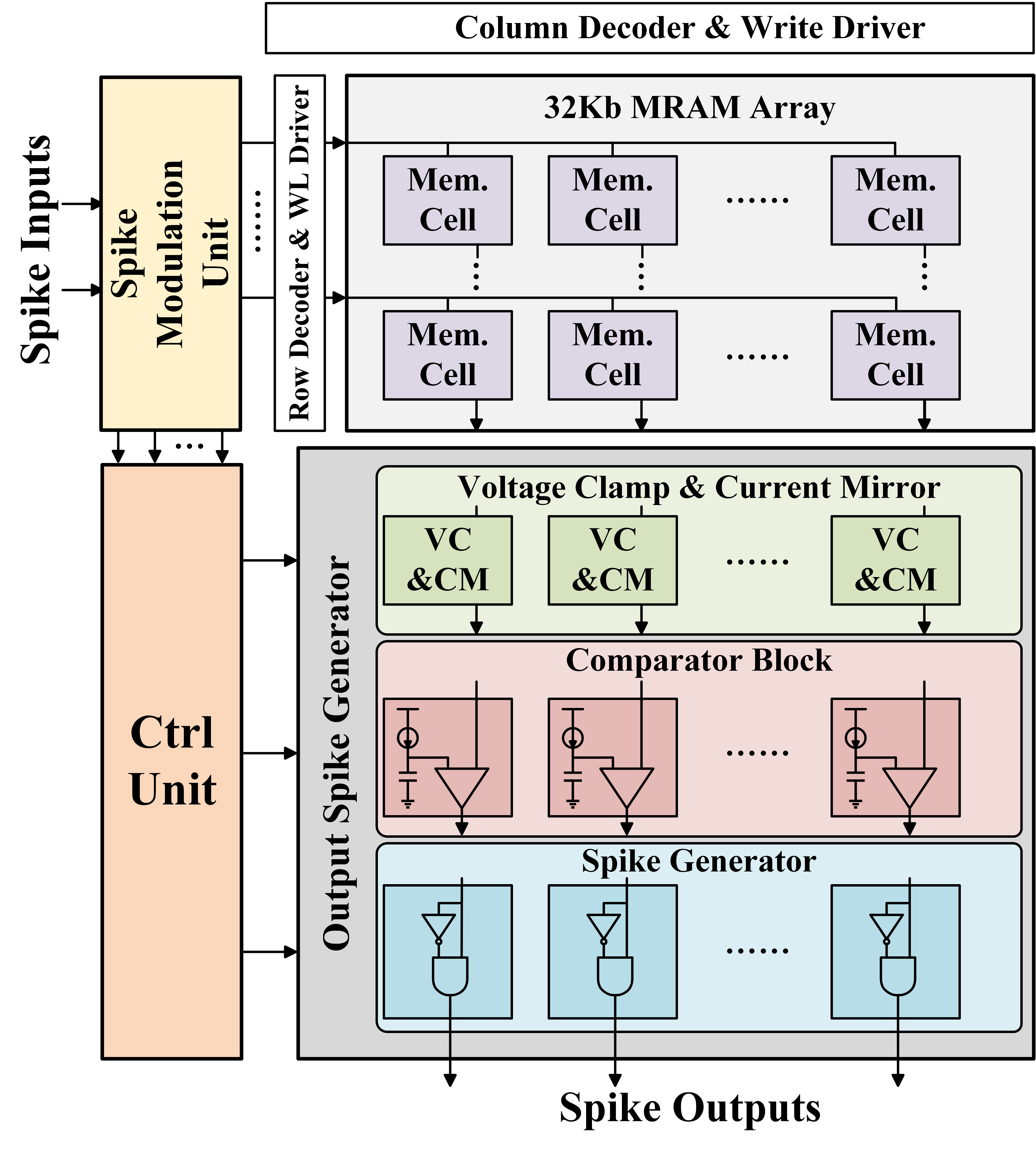}
\vspace{-4mm}
\caption{Overall architecture of the proposed spike-based MRAM CIM macro.}
\label{fig:overall-arch}
\end{center}
\end{figure}

\subsection{Related Works on Spiking Processing}\label{2B}


To eliminate the need for power-hungry ADCs and DACs, \cite{18,21} proposed  rate-coded architectures, demonstrating the feasibility of spiking computation.
However, these designs still suffer from limited accuracy and suboptimal energy efficiency.
Beyond rate coding, temporal coding schemes such as TTFS and dual-spike coding have been explored. 
Despite their potential, existing designs face critical challenges, including degraded computational accuracy due to the nonlinearity of output sensing circuits~\cite{22,23}, and reliance on global clock synchronization, which undermines the inherent benefits of spiking computation~\cite{14,24}.
In this work, we adopt a dual-spike coding scheme and propose novel circuit and system level designs that enable event-driven processing with superior energy efficiency and high-accuracy computation ensured by linear output sensing.


\section{The Proposed SOT-MRAM Macro}
Fig.~\ref{fig:overall-arch} illustrates the overall architecture of the proposed 32Kb spike-based SOT-MRAM CIM macro for analog MVM operations.
The macro comprises three key components: a 3T-2MTJ SOT-MRAM crossbar array, a spike modulation unit that converts input spikes into stable read voltages, and an output spike generator that transforms the resulting current into linearly encoded output spikes.


\begin{figure}[b]
\centerline{\includegraphics[width=\columnwidth]{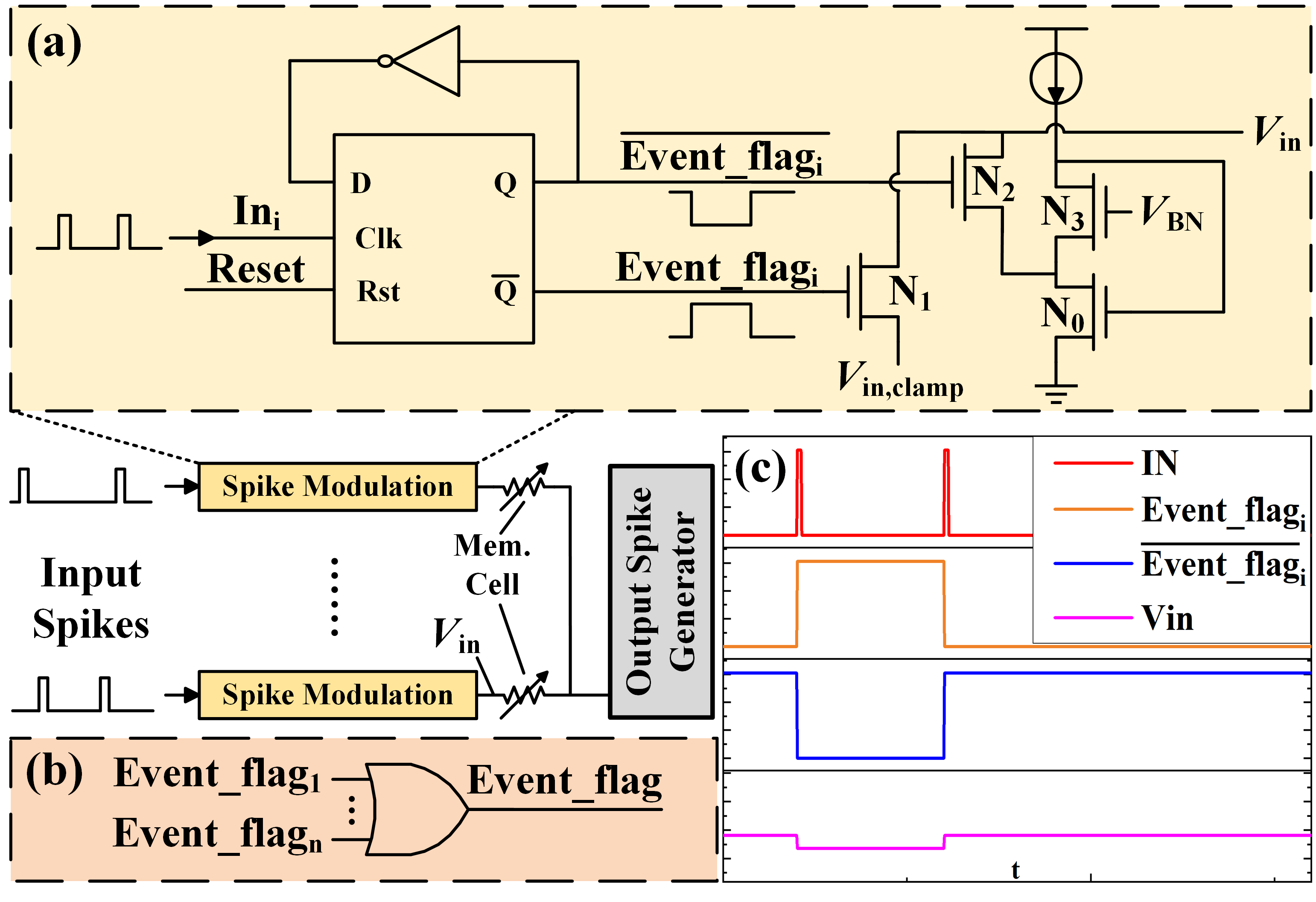}}
\caption{(a) Schematic of spike modulation unit; (b) Formation of Event\_sig; (c) Transient simulation of spike modulation unit.}
\label{fig:spike}
\end{figure}

\subsection{3T-2MTJ Memory Cell}


The proposed SOT-MRAM array comprises 128$\times$128 memory cells, each implemented with a three-transistor and two-SOT-MRAM (3T-2MTJ) structure. 
As shown in Fig.~\ref{fig:Traditional} (b), during write operations, all transistors are activated, allowing currents to pass through the heavy-metal layer of the two SOT-MRAMs and switch the magnetization state of the MTJs.
For read operations, the transistors remain off, and the cell presents a combined resistance determined solely by the two MRAMs.
To support multi-bit storage, \ensuremath{J_2} is designed with twice the resistance of \ensuremath{J_1}, yielding four distinct resistance states that encode 2-bit data.
Within each column, the RBL[1] lines of all cells are connected to the readout circuit, while RBL[0] serves as the input line.
MVM is performed in parallel by applying dual-spike input signals across all 128 rows simultaneously.


\begin{figure}[t]
\centerline{\includegraphics[width=0.95\columnwidth]{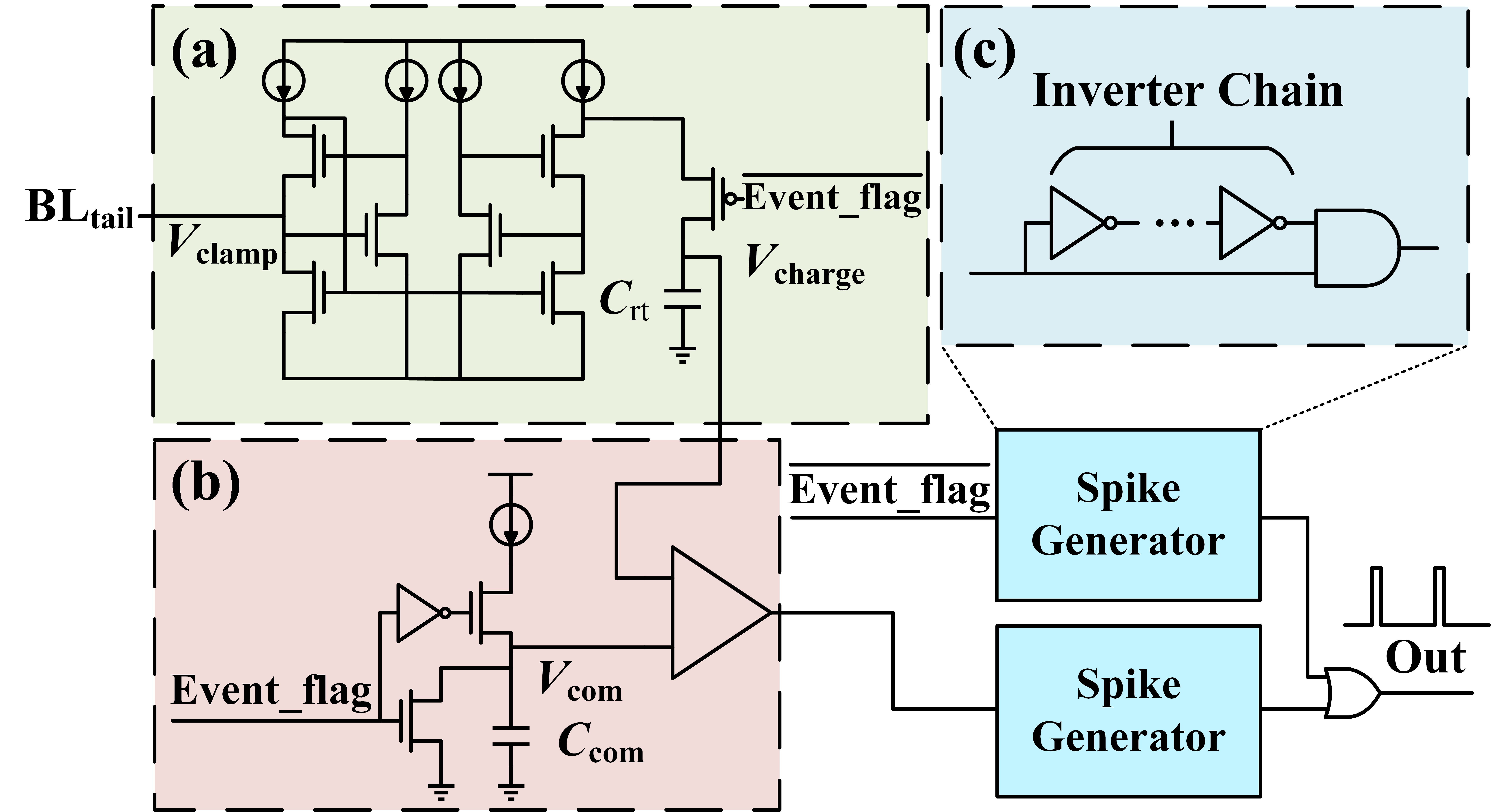}}
\caption{Schematics of output spike generator: (a) clamping\&current mirror; (b) comparator block; (c) spike generator.}
\label{fig:OUT}
\end{figure}

\subsection{Spike Modulation Unit}
\label{sec:smu}
We design a spike modulation unit (SMU) that converts double-spike inputs into a stable read voltage, with its duration determined by the inter-spike interval.

As shown in Fig.~\ref{fig:spike} (a), the SMU consists of a D flip-flop (DFF) and an input clamping circuit. 
The DFF generates a flag pulse \(Event\_flag_{\text{i}}\), encoding the input event duration and serving as the foundation for the event-driven computation.
All \(Event\_flag_{\text{i}}\) signals are then aggregated to produce a global \(Event\_flag\) (Fig.~\ref{fig:spike}  (b)), enabling system-level asynchronous control (detailed in Section~\ref{sec:cog}).
The input clamping circuit is designed to regulate the input voltage \( V_{\text{in}} \) of the crossbar array based on \( Event\_flag_{\text{i}} \): when high, it activates transistor \ensuremath{N_1}\ to pull \( V_{\text{in}} \) to \( V_{\text{in, clamp}} \); when low, it activates \ensuremath{N_2}\ to set \( V_{\text{in}} \) to \( V_{\text{clamp}} \).
This ensures that a fixed read voltage \( V_{\text{read}} \)$=$\( V_{\text{clamp}} - V_{\text{in,clamp}} \) is applied to the memory cell only during active spike events.
Fig.~\ref{fig:spike} (c) shows the simulated waveforms of the SMU, where the generated \( Event\_flag_{\text{i}} \) stabilizes \( V_{\text{in}} \) in accordance with spike timing. 
\subsection{Output Spike Generator}
\label{sec:cog}

The output spike generator (OSG) converts the MVM current result from each column of the SOT-MRAM crossbar into spiking outputs. 
As shown in Fig.~\ref{fig:OUT}, the OSG comprises three modules: a clamping and current-mirror (Clamping\&CM) circuit, a comparator, and a spike generator. 
As mentioned above, the clamping circuit fixes the \( \text{RBL}[1] \) terminal of each memory cell to a fixed voltage \( V_{\text{clamp}} \). 
The accumulated column current is mirrored to charge a capacitor \( C_{\text{rt}} \), as depicted in Fig.~\ref{fig:OUT} (a). 
This current mirror structure is essential to preserve computational linearity, as directly charging the \( C_{\text{rt}} \) with the bit-line current would result in exponential voltage growth. 
The capacitor charging duration is controlled by \( Event\_flag \), which de-asserts once all input events complete.  
At that point, the comparator and spike generator modules are triggered to produce the corresponding output spike.

\begin{figure}[t]
\centerline{\includegraphics[width=0.8\columnwidth]{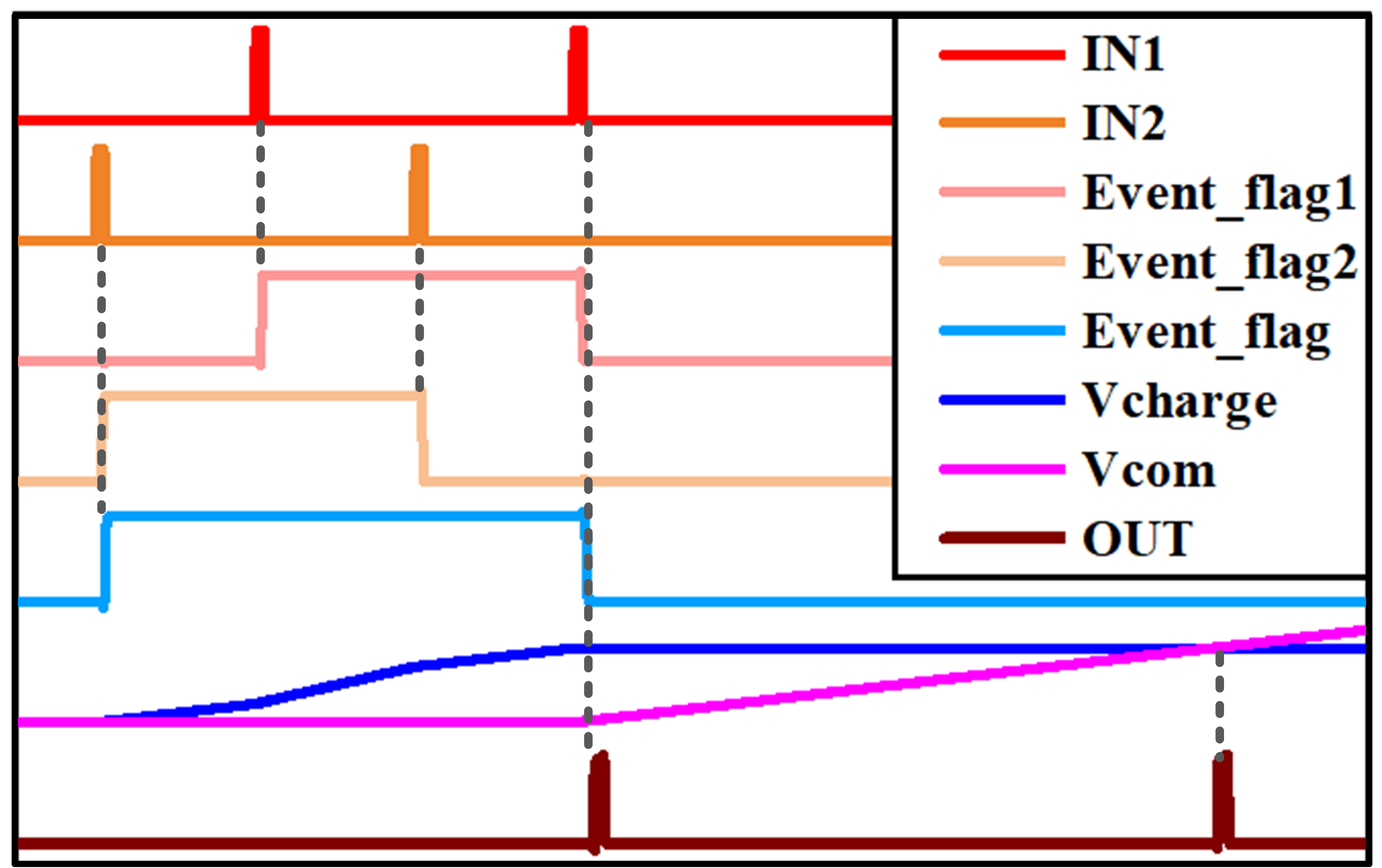}}
\caption{Transient simulation results of proposed design.}
\label{fig:CURVES}
\end{figure}

The spike generator emits a spike in response to rising input edges. 
As shown in Fig.~\ref{fig:OUT} (c), the first spike of the output spike pair is triggered by the rising edge of \( \overline{Event\_flag} \).
Meanwhile, a reference capacitor \( C_{\text{com}} \) begins charging at a fixed current \( I_{\text{com}} \) under \( {Event\_flag} \) control (Fig.~\ref{fig:OUT} (b)). 
The voltages across \( C_{\text{rt}} \) and \( C_{\text{com}} \), i.e., \( V_{\text{charge}} \) and \( V_{\text{com}} \) are continuously monitored by a comparator. 
When \( V_{\text{com}} \) reaches \( V_{\text{charge}} \), the comparator toggles, and its output rising edge is captured by a second spike generator, which emits the second spike in the output pair. 
Consequently, the time interval between the two output spikes precisely encodes the analog MVM current result, completing the spike-based sensing operation. 
The simulation results in Fig.~\ref{fig:CURVES} validate this behavior.
Theoretically, the following equation is satisfied when the second output spike is generated:
\begin{equation}
kC_{\mathrm{rt}} \sum_i T_{\mathrm{in},i} V_{\mathrm{read}} G_{\mathrm{mem},i} = C_{\mathrm{com}} I_{\mathrm{com}} T_{\mathrm{out}}
\end{equation}

\begin{figure}[b]
\begin{center}
\hspace{-2mm}
\includegraphics[width=0.95\columnwidth]{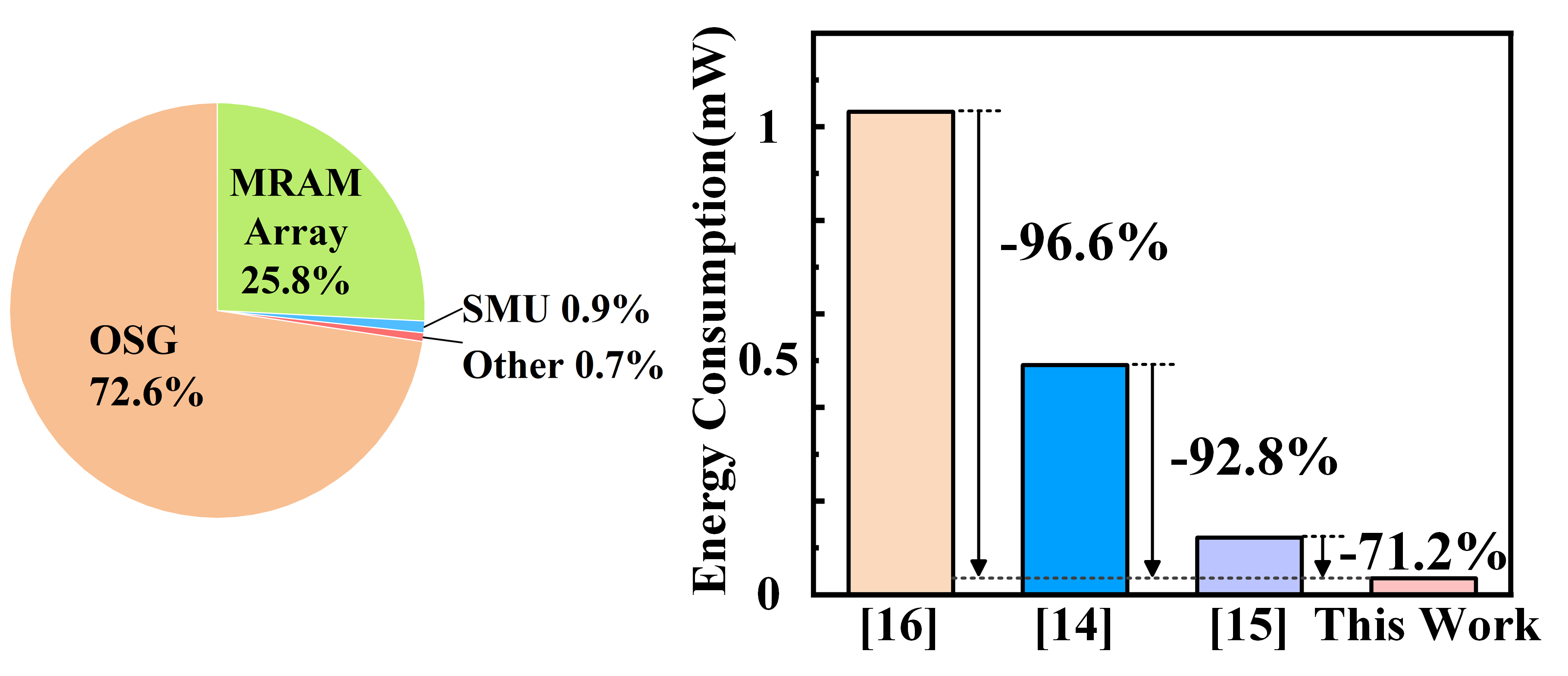}
\caption{(a) Power breakdown of whole design; (b) Energy consumption comparison of sensing circuits.}
\label{fig:PowerBreakdown}
\end{center}
\end{figure}
Here, \(k\) is the current-scaling factor of the current mirror, \( T_{\text{in,$i$}} \) denotes the time interval between the two input spikes applied to the $i$th memory cell in each column, and \( G_{\text{mem,$i$}} \) represents the conductance of the $i$th memory cell. 
Therefore, the time interval between the two output spikes \( T_{\text{out}} \) can be expressed as:
\begin{equation}
T_{\mathrm{out}} = \alpha\sum_i T_{\mathrm{in},i} G_{\mathrm{mem},i}
\end{equation}
where $\alpha = kV_\mathrm{read}C_\mathrm{rt} / (I_\mathrm{com}C_\mathrm{com})$
is a constant determined by the aforementioned circuit parameters.
This indicates that the proposed sensing circuit establishes a linear relationship between \( T_{\text{out}} \) and $\sum_i T_{\text{in},i} G_{\text{mem},i}$, thus ensuring the computational accuracy of our spiking design.


\section{Evaluation}

We implement and evaluate the proposed MRAM CIM macro using Cadence Virtuoso under a 28 nm CMOS technology node.
The key parameters of the simulation are summarized in Table~\ref{tab:sim_parameters}.
Each memory array consists of a 128$\times$128 3T-2MTJ SOT-MRAM cells. 
In the simulation, a time interval of 0.2ns was assigned to represent each bit. The charging and comparison capacitors \( C_{\text{rt}} \) and \( C_{\text{com}} \) are both set to be 200fF.
\begin{table}[H]
\centering
\caption{Key Parameters of Simulation}
\renewcommand{\arraystretch}{1.3}
\Large
\resizebox{0.95\linewidth}{!}{
\begin{tabular}{|c|c|c|c|c|}
\hline
\textbf{Parameters} & Cell Structure & Supply Voltage & $\text{R}_{\text{LRS}}$ of MTJ & TMR \\
\hline
\textbf{Value} & 3T-2J & 1.1V & 1M$\Omega$~\cite{26} & 100\% \\
\hline
\end{tabular}
}
\label{tab:sim_parameters}
\end{table}
\noindent In the spike modulation unit, the clamping voltage \( V_{\text{in,clamp}} \) is set to 300mV, while \( V_{\text{clamp}} \) is 400mV, resulting in an effective read voltage \( V_{\text{read}} \) of approximately 100mV.

\begin{table*}[!t]
\centering
\caption{Comparison with Other CIM Designs}
{\fontsize{8.5pt}{10pt}\selectfont   
\setlength{\tabcolsep}{10pt}   
\begin{tabular}{|c|c|c|c|c|c|c|c|}
\hline
Work & VLSI’19~\cite{18} & DAC’20~\cite{14} & TCAS-I’22~\cite{24} & ESSCIRC’21~\cite{13} & DAC’24~\cite{16} & \textbf{This Work} \\
\hline
Memory Type & ReRAM & ReRAM & ReRAM & MRAM & MRAM & \textbf{MRAM} \\
Technology Node & 150nm & 65nm & 65nm & 22nm & 28nm & \textbf{28nm} \\
Cell Structure & 1T-1R & 1T-1R & 1T-1J & 2T-2J & 6T-4J & \textbf{3T-2J} \\
Array Size & 256$\times$256 & 32$\times$32 & 128$\times$128 & 128$\times$128 & 64$\times$128 & \textbf{128$\times$128} \\
Readout Scheme & CA+IFC & COG & LIF & ADC & ADC & \textbf{OSG} \\
Efficiency (TOPS/W) & 16.9 & 40.8 & 46.6 & 5.1 & 23.7-29.4 & \textbf{243.6} \\
\hline
\end{tabular}
}
\vspace{-2mm}
\label{tab:PIMcomparison}
\end{table*}

\subsection{Energy Efficiency Analysis and Comparison}

\begin{figure}[b]
\begin{center}
\hspace{-2mm}
\includegraphics[width=0.9\columnwidth]{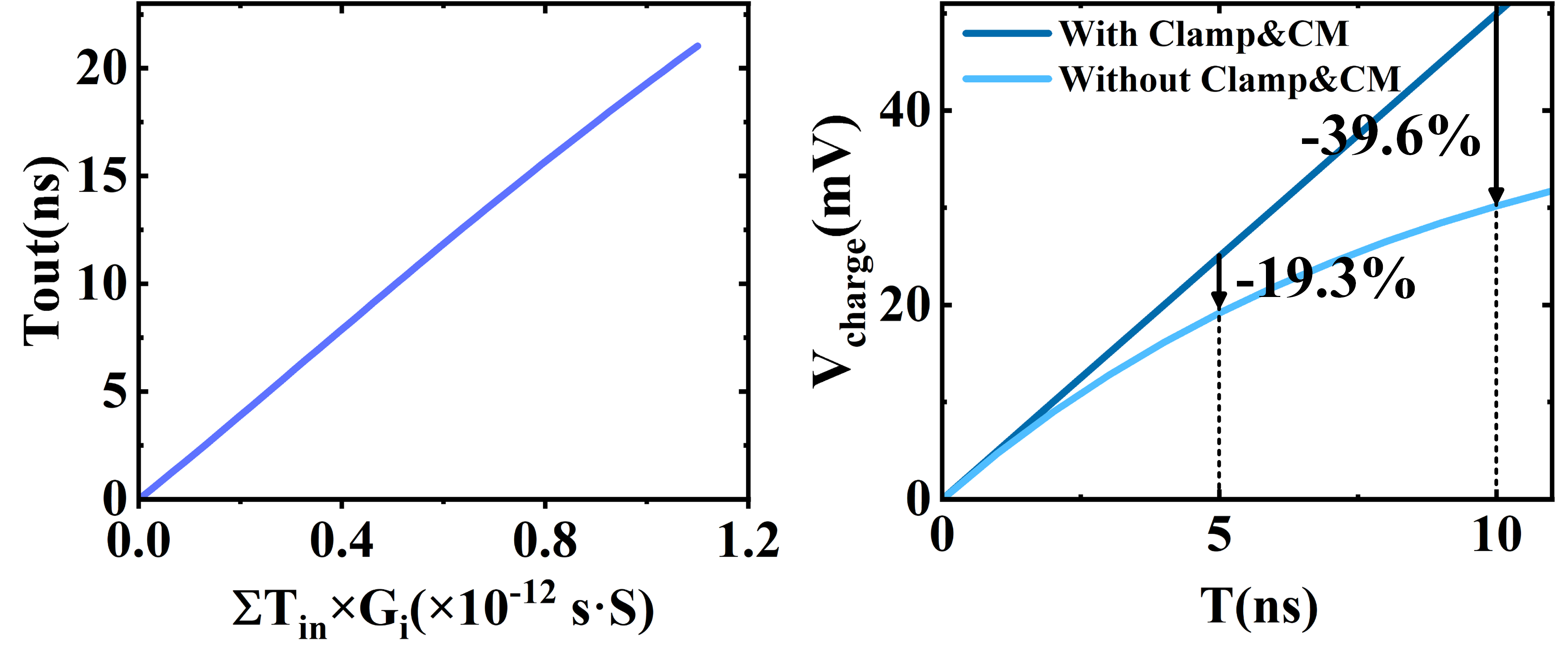}
\caption{(a) MAC computation result \( T_{\text{out}} \) vs. $\sum_i T_{\text{in},i} G_{\text{mem},i}$; (b) \( V_{\text{charge}} \) comparison with and without clamp\&current mirror structure.}
\label{fig:Linearity}
\end{center}
\end{figure}

Fig.~\ref{fig:PowerBreakdown} (a) illustrates the power consumption breakdown of the proposed SOT-MRAM macro. 
The output spike generator dominates the total power consumption, accounting for 72.6\% of the overall budget. 
In our design, MRAM devices with high resistance values (M$\Omega$ level) are employed, which naturally contribute to improving the overall energy efficiency. 
Therefore, to ensure a fair comparison, we further evaluate the power consumption of the readout and sensing circuits between our design and prior works, including ADC-based~\cite{16}, TDC-based~\cite{15}, and custom spike-based implementations~\cite{14}, as shown in Fig.~\ref{fig:PowerBreakdown} (b). Specifically, the proposed design achieves energy reductions of 96. 6\%, 92. 8\% and 71. 2\% compared to ~\cite{16}, ~\cite{14} and ~\cite{15}, respectively.

We further compare the energy efficiency of our proposed SOT-MRAM-based CIM macro with several representative CIM designs, as summarized in Table~\ref{tab:PIMcomparison}. 
The selected baselines cover a variety of data representation schemes, including rate-coded designs~\cite{18}, clock-dependent temporal coding~\cite{14,24}, and analog-voltage-based architectures that rely heavily on ADCs~\cite{13,16}. Experimental results show that the proposed CIM macro achieves an energy efficiency of 243.6 TOPS/W under 8-bit input data precision.
This significant improvement is mainly attributed to two key innovations: 1) an event-driven spike processing paradigm, which activates the memory array only upon spike events, thereby significantly reducing idle power consumption; 2) lightweight read out and sensing circuitry that eliminate energy-hungry analog components, such as ADCs.

\subsection{Computing Linearity and Accuracy}

In our design, the output spike interval \( T_{\text{out}} \) is linearly proportional to the weighted sum of the memory-cell conductance and the corresponding input spike intervals, as shown in Eq. (2). 
To evaluate the computational linearity of the proposed architecture, a large number of 8-bit input data and 2-bit weight combinations were applied, uniformly distributed across the entire input–weight space.
The relationship between \( T_{\text{out}} \) and $\sum_i T_{\text{in},i} G_{\text{mem},i}$ is illustrated in Fig.~\ref{fig:Linearity}(a), demonstrating excellent linearity.
This confirms that the proposed spike-based CIM macro can perform the analog MVM operations in extremely high accuracy.

To further validate the necessity of the Clamping\&CM circuit, we compare the voltage accumulation on the result capacitor \( C_{\text{rt}} \) charged by the bit line current under two configurations: with and without this circuit. As shown in Fig.~\ref{fig:Linearity}(b), without the Clamping\&CM circuit, the accumulated voltage \( V_{\text{charge}} \) exhibits a noticeable drop due to current degradation. 
In contrast, incorporating of the Clamping\&CM circuit maintains a linear charging profile, effectively mitigating voltage loss and enhancing sensing accuracy. 
Quantitatively, the \( V_{\text{charge}} \) degradation reaches 19.3\% at a 5ns charging duration and increases to 39.6\% at 10ns. 
As high bit data precision in our design requires longer charging periods, this circuit plays a crucial role in ensuring accurate computation results, compared to previous approaches in which the result capacitor is directly charged by the bit line current~\cite{14,15,23}.

\section{Conclusion}
In this work, we propose an event-driven spiking CIM macro based on SOT-MRAM, which addresses key challenges in existing designs through a spike-triggered processing scheme and high-efficiency, high-linearity readout circuitry.
The CIM macro has three key components -- a 3T-2MTJ SOT-MRAM crossbar array, a spike modulation unit, and an output spike generator.
Implemented on a 128$\times$128 crossbar under 28nm technology, the proposed design achieves robust computational accuracy and excellent energy efficiency.
The proposed readout and sensing circuitry achieves up to 96.6\% energy saving, and the overall system delivers 243.6TOPS/W under 8-bit input precision, significantly outperforming state-of-the-art designs.
These results underscore the potential of spike-based SOT-MRAM CIM architectures as a promising solution for next-generation low-power neural network accelerators, and establish a solid foundation for scalable analog computing in future intelligent systems.

\end{document}